# Distributed-element circuit model of edge magnetoplasmon transport

Masayuki Hashisaka[1,*], Hiroshi Kamata[1], Norio Kumada[2], Kazuhisa Washio[1], Ryuji Murata[1],
Koji Muraki[2], Toshimasa Fujisawa[1]

[1]*Department of Physics, Tokyo Institute of Technology, 2-12-1-H81 Ookayama, Meguro, Tokyo 152-8551, Japan*
[2]*NTT Basic Research Laboratories, NTT Corporation, 3-1 Morinosato-Wakamiya, Atsugi, Kanagawa 243-0198, Japan*
[*]*e-mail: hashisaka@phys.titech.ac.jp, Phone: +81-3-5734-2809*

We report experimental and theoretical studies of edge magnetoplasmon (EMP) transport in quantum Hall (QH) devices. We develop a model that allows us to calculate the transport coefficients of EMPs in QH devices with various geometries. In our model, a QH system is described as a chiral distributed-element (CDE) circuit, where the effects of Coulomb interaction are represented by an electrochemical capacitance distributed along unidirectional transmission lines. We measure the EMP transport coefficients through single- and coupled-edge channels, a quantum point contact, and single- and double-cavity structures. These measured transmission spectra can be reproduced well by simulations using the corresponding CDE circuits. By fitting the experimental results with the simulations, we deduce the circuit parameters that characterize the electrostatic environment around the edge channels in a realistic QH system. The observed gate-voltage dependences of the EMP transport properties in gate-defined structures are explained in terms of the gate tuning of the circuit parameters in CDE circuits.

## I. INTRODUCTION

The application of radio-frequency (rf) electronic signals to a one-dimensional (1D) quantum Hall (QH) edge state excites charge density waves, called edge magnetoplasmons (EMPs) [1-10], that travel chirally along the edge channels. Because EMPs travel coherently over more than a few millimeters, it is feasible to design coherent plasmonic circuits [11], cavities [3-6], and crystals [11,12] for EMPs. Such plasmonic devices can be functionalized by exploiting various electrostatic tuning techniques using surface metal gates. As in typical dc transport experiments, one can switch the paths of EMP transport by selectively activating metal gates to deplete the two-dimensional electron gas (2DEG) underneath. More importantly, gate-defined structures provide further novel functionality to plasmonic devices via the sensitivity of EMPs to their electrostatic environment. When the channel is covered with a metal, the group velocity of EMPs ($v_\text{EMP}$) along an etched QH boundary (bare edge channel) is significantly reduced (by more than one order of magnitude) because of the screening of the electric field [6-8,10]. As compared to bare edge channels, those defined with gates (gate-defined edge channels) have the advantage that $v_\text{EMP}$ is tunable, because the screening effect can be tuned by changing the distance between the channel and the metal electrode via a gate voltage [9]. This feature can be exploited to form tunable delay lines in plasmonic circuits. In addition, a beam splitter for EMPs can be constructed from a quantum point contact (QPC), whose transmission and reflection coefficients can be changed electrostatically [13-16]. Frequency multipliers and mixers based on the nonlinearity of the transmission characteristics of a QPC have been demonstrated [17]. In this way, we can expect to establish integrated plasmonic circuits by combining such EMP devices with electrostatic tuning techniques.

To design and prepare an appropriate electrostatic environment for EMP transport, it is essential to correctly evaluate the effect of Coulomb interaction arising from the non-uniform charge density modulation of EMPs. However, the long-range nature of the Coulomb interaction makes the calculation cumbersome, particularly when multiple plasmonic devices are integrated on a chip. A tractable model with a certain accuracy is necessary to deal with EMP transport in actual devices. In this paper, we propose a simple method that allows us to describe the effect of Coulomb interaction and dissipation on EMP transport in terms of a distributed-circuit model. In this model, the wave equation for EMPs traveling along a single edge channel is expressed in terms of the distributed electrochemical capacitance between the channel and the ground ($c_\text{ch}$: channel capacitance). In addition, electrostatic coupling between two different channels can be modeled with distributed elements, which is expressed by the inter-edge capacitance ($c_\text{X}$: inter-edge capacitance) [16]. Once the transport characteristics of each individual device are obtained with the relevant distributed capacitance, one can simulate EMP transport in an integrated system by using the scattering matrix formalism. Since the above modeling method is developed for chiral edge channels, we call it "the chiral distributed-element (CDE) circuit model."

We apply this CDE model to EMP resonance in single and double plasmonic cavities and compare the results of the calculations with those of experiments. Our device is designed such that a single or double cavity can be selectively activated. As we shall see, the full set of cavity parameters can be controlled electrostatically, including the resonant frequency, coupling between the cavities, and coupling to the outside channels. This allows us to study several types of devices with different characteristics in one sample. We start from the simplest device, a QPC beam splitter, and then investigate single and coupled cavities. The measured characteristics of all these devices show reasonable agreement with the calculations for the corresponding CDE circuits. In particular, our experiment successfully identifies the in-phase and anti-phase coupling modes in a coupled cavity predicted by the calculation. These findings ensure the validity and usefulness of the CDE model for developing future plasmonic circuits.

The rest of this paper is organized as follows. In section II, a detailed account of the CDE model is given. We start by developing a model for a single edge channel, and then extend it to other structures: coupled edge channels, a QPC, and single and coupled cavities. In section III, we present experimental results and compare them with the calculations based on the CDE model. In section IV, we discuss the scope of the model's application and present a few prospects and comments regarding improving the model and designing future plasmonic circuits. Section V is devoted to the conclusion.

## II. CHIRAL DISTRIBUTED-ELEMENT CIRCUIT

The transport properties of an EMP, including its dispersion relation, can in principle be derived by solving the equation of motion and the continuity equation of charge at the edge of the QH system, together with the Poisson equation, which relates the excess charge of EMPs to the electric field [1,2]. Such a hydrodynamic approach, however, is not readily applicable to real devices, where the Poisson equation must be solved in three dimensions. In this section, we give a detailed account of the CDE model, which enables us to simulate EMP transport in actual QH devices. All the essential ingredients of EMP transport, such as the wave equations along the edge channels and the scattering matrices, are naturally derived from its circuit representation.

We consider a 2DEG with electron density $n_e$ placed in a high magnetic field $B$ perpendicular to it. Its dc transport characteristics are described by the longitudinal conductivity $\sigma_{xx}$ and Hall conductivity $\sigma_{xy} \approx en_e/B$. We restrict our discussion to the situation $|\sigma_{xx}| \ll |\sigma_{xy}|$, i.e., the case where EMP transport is dominant and bulk plasmons are negligible. For simplicity, in the following argument we discuss only the fundamental EMP mode and neglect the acoustic mode [2].

### A. Unidirectional transmission line

In our circuit representation, an edge channel is modeled as a unidirectional transmission line consisting of a channel and a ground, as shown schematically in Fig. 1(a). We take the $x$ axis to be parallel to the transmission line. We consider the relation between the voltage $V(x, t)$ and current $I(x, t)$ flowing in the channel at position $x$ and time $t$. $V(x, t)$ is related to the excess charge distribution $\rho(x, t)$ as $\rho = c_{ch}V$ through the channel capacitance $c_{ch}$, which represents the effective electrochemical capacitance (per unit length) between the channel and the ground. On the other hand, $I(x, t)$ is related to $\rho(x, t)$ as $\partial I/\partial x = -\partial \rho/\partial t$ according to the continuity equation along the 1D channel. Here, we assume that $I(x, t)$ and $V(x, t)$ satisfy the relation $I(x, t) = \sigma_{xy}V(x, t)$ [18]. In terms of circuitry, this implies that we can define the characteristic impedance of the transmission line as $Z_+ = V/I = 1/\sigma_{xy}$. From these three relations, we obtain the wave equation for EMPs as

$$\frac{\partial I(x,t)}{\partial t} = -\frac{\sigma_{xy}}{c_{ch}}\frac{\partial I(x,t)}{\partial x}. \quad (1)$$

The general solution of this equation takes the form $I(x - v_{EMP}t)$ with velocity

$$v_{EMP} = \frac{\sigma_{xy}}{c_{ch}}, \quad (2)$$

whereas the form $I(x + v_{EMP}t)$ cannot fulfill eq. (1). Note that the direction of propagation is determined by the sign of $\sigma_{xy}$, that is, by the sign of $B$. This unidirectional nature stems from the fact that eq. (1) is expressed by the first derivative with respect to both space and time, and corresponds to the chiral transport nature of EMPs. In the circuit representation of the transmission line of Fig. 1(a), $c_{ch}$ is expressed as the distributed admittance $Y_p = i\omega c_{ch}$ for generality. The chirality of the channel is marked by arrows.

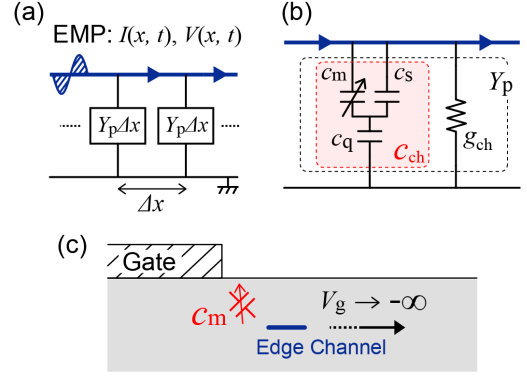

Fig. 1 (a) Circuit representation of an EMP transmission line. (b) Components of the admittance $Y_p$ between the channel and the environmental ground. (c) Schematic of the screening capacitance $c_m$ of a gate-defined channel. The distance between the channel and the metal, and hence $c_m$, can be tuned with the gate voltage $V_g$.

Let us consider the channel capacitance $c_{ch}$ in the circuit representation. In mesoscopic systems, the electrochemical capacitance is described as a series of the electrostatic geometrical capacitance and quantum capacitance $c_q \equiv e^2D(E)$, where $D(E)$ is the electron density of states [13,19-21]. In a 1D system, $c_q = e^2D(E) = e^2/hv_F$, where $v_F$ is the Fermi velocity. Further, $v_F$ corresponds to the mean drift velocity $v_D = (dU/dy)/B$ due to the Lorentz drift in an electric field $dU/dy$, which is determined by the gradient of the edge confinement potential $U$ at the boundary of the 2DEG [20]. First, we consider the strong-screening limit, where Coulomb interaction between electrons can be ignored because of the screening of the electric field provided by the surrounding metals. In this limit, the electrostatic geometrical capacitance is much larger than $c_q$. Hence, $c_{ch}$ is governed by $c_q$ in the series connection; i.e., $c_{ch} \cong c_q$. This leads to $v_{EMP} \cong v_D$ from eq. (2); electrons travel as non-interacting Fermions with the drift velocity $v_D$ rather than as plasmons in this strong-screening limit. For example, for a smooth confinement potential $U$, one finds $c_q = e/(2\pi l_m^2|dU/dy|) \cong 5$ nF/m and $v_D \cong 10^4$ m/s for a typical GaAs heterostructure at $B = 5$ T, where $l_m$ is the magnetic length [22].

In actual QH conductors, however, the screening effect is not strong enough for this limit to be realized. Usually, $v_{EMP}$ is about two orders of magnitude higher than $v_D$ ($v_{EMP} \cong 10^6$ m/s in a filling factor $\nu = 1$ QH state [5-7,9,10]). This is a manifestation of the strong Coulomb interaction in a bare edge channel. We include the effect of this Coulomb interaction by using the electrostatic capacitance. In principle, the long-range nature of the Coulomb interaction makes the problem essentially non-local, which implies that $V(x)$ should depend on $\rho(x')$ at any position $x'$ in the channel. For an infinitely long straight edge channel, however, because of the translational symmetry of the system, the simple relation $\rho(x) = c_{ch}(k)V(x)$ holds for a given wavenumber $k$. The effect of non-local $\rho(x')$ is incorporated as the $k$ dependence of $c_{ch}(k)$ in this relation. Indeed, the calculation of $v_{EMP}$ in such a scheme shows a similar $k$-dependence of $v_{EMP}$ = $\sigma_{xy}/\varepsilon \times \ln(\gamma/|ka|)$ for an abrupt electron density profile [1] and a soft profile [2] at the QH boundary. Here, $\varepsilon$ is the dielectric constant, $a$ is the effective width of the edge channel, and $\gamma$ is a numerical constant. Comparing this expression for $v_{EMP}$ with eq. (2), one can define a $k$-dependent effective self-capacitance $c_s(k)$ = $\varepsilon/\ln(\gamma/|ka|)$. Its weak logarithmic dependence can be neglected for a certain wavelength region of interest, where $c_s$ is expressed by considering $c_s = \varepsilon/\gamma^*$; $\gamma^* = \ln(\gamma/|k_{typ}a|)$ is a constant, and $k_{typ}$ is the typical wavenumber. Because $c_s$ is much smaller than $c_q$ in a typical GaAs heterostructure, $v_{EMP} \cong \sigma_{xy}/c_s$ is generally much higher than $v_{EMP} \cong v_D$.

In a realistic QH device, in which such exact translational symmetry does not exist, the charge distribution $\rho(x')$ along the entire channel must be taken into account. When the device is covered with a metal electrode, however, partial screening of the long-range Coulomb interaction allows us to use a local approximation that considers the Coulomb interaction only near the channel. This is done by replacing $c_s$ with the electrostatic screening capacitance $c_m = \varepsilon a/d$ between the channel and the metal, where $d$ is the distance between the channel and the covering metal. In this case, $v_{EMP}$ is given by $v_{EMP} \cong \sigma_{xy}/c_m \cong \sigma_{xy}d/\varepsilon a$ for $d \ll a$ [7,8,10]. Generally, by considering all the contributions of $c_q$, $c_s$, and $c_m$, the channel capacitance $c_{ch}$ is conceptually understood using an equivalent circuit, as shown in Fig. 1(b). The electrostatic screening capacitances $c_s$ and $c_m$ are connected to $c_q$ in series. Because $c_m$ for a metal-covered channel is generally much larger than $c_q$ and $c_s$, the channel capacitance $c_{ch}$ is approximated as $c_{ch} \cong c_m$.

This capacitance approximation also adequately explains the behavior of $v_{EMP}$ along a gate-defined QH boundary. The capacitance $c_m$, and hence $v_{EMP}$, can be tuned with a negative gate voltage $V_g$ because the distance between the channel and the metal varies with $V_g$ [Fig. 1(c)] [9]. In this way, the variation of $v_{EMP}$ along gate-defined channels is represented as the variation of a circuit parameter $c_m$. The velocity $v_{EMP}$ can be estimated by numerically calculating $c_m$ for a realistic electron density profile at the gate-defined QH boundary.

The longitudinal conductance $\sigma_{xx}$, which was ignored in the above discussion, causes current leakage from the channel into the bulk region and the resultant dissipation of EMPs [1]. This source of EMP dissipation can be represented by introducing a finite conductance $g_{ch}$ between the channel and the ground in the CDE model as shown in Fig. 1(b). The distributed admittance $Y_p$ depicted in Fig. 1(a) is then described by $Y_p = i\omega c_{ch} + g_{ch}$.

**B. Coupled transmission lines**

In this subsection, we formulate EMP transport in a pair of parallel edge channels in order to extend the CDE model to cases where one must consider Coulomb interaction between edge channels. The inter-edge Coulomb interaction is a key factor in understanding the electron dynamics in QH edge channels, and is believed to dictate the energy relaxation [23,24] and charge fractionalization [25]. Since our CDE model allows us to parameterize and evaluate the inter-edge Coulomb interaction, it will be helpful for studying the non-trivial electron dynamics that emerges in coupled 1D systems.

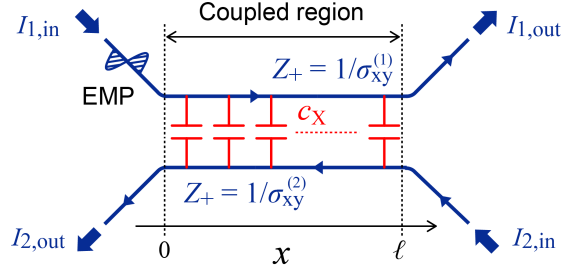

Fig.2 Schematic of coupled transmission lines. Two unidirectional transmission lines are coupled for length $l$ through inter-edge capacitance $c_X$.

Figure 2 shows our model for a coupled system in which two counter-propagating edge channels interact over a finite length $l$ through inter-edge capacitance $c_X$. We take the $x$ axis to be parallel to the edge channels in the interacting region. We assume that the two unidirectional transmission lines $i$ ($i$ = 1 for the upper channel and 2 for the lower channel) have a characteristic impedance $1/\sigma_{xy}^{(i)}$, where $\sigma_{xy}^{(1)} = -\sigma_{xy}^{(2)} = \sigma_{xy}$, and identical channel capacitances $c_{ch}^{(1)} = c_{ch}^{(2)} = c_{ch}$. In the coupling region ($0 < x < l$), the charge density $\rho_i(x)$ and potential $V_i(x)$ ($i$ = 1, 2) of the two channels are interrelated as in the following matrix form

$$\begin{pmatrix} \rho_1 \\ \rho_2 \end{pmatrix} = \begin{pmatrix} c_{ch} + c_X & -c_X \\ -c_X & c_{ch} + c_X \end{pmatrix} \begin{pmatrix} V_1 \\ V_2 \end{pmatrix}. \quad (3)$$

On the other hand, the relation between the potential $V_i$ and current $I_i$ for each channel $i$ is $I_i = \sigma_{xy}^{(i)} V_i$. From these relations, along with the continuity equation $\partial I_i/\partial x = -\partial \rho_i/\partial t$, the wave equation for the coupled system is

$$\frac{\partial}{\partial t}\begin{pmatrix} I_1 \\ I_2 \end{pmatrix} = -\begin{pmatrix} U_0 & U_{12} \\ -U_{12} & -U_0 \end{pmatrix} \frac{\partial}{\partial x}\begin{pmatrix} I_1 \\ I_2 \end{pmatrix}, \quad (4)$$

where $U_0$ and $U_{12}$ are the intra- and inter-channel coupling, respectively:

$$U_0 = \frac{\sigma_{xy}}{c_{ch}} \frac{c_{ch} + c_X}{c_{ch} + 2c_X},$$

$$U_{12} = \frac{\sigma_{xy}}{c_{ch}} \frac{c_X}{c_{ch} + 2c_X}.$$

The above wave equation has two solutions, one that fulfills $I_1 = -rI_2$ and the other that satisfies $I_2 = -rI_1$, with $r = (U_0 - \sqrt{U_0^2 - U_{12}^2})/U_{12}$. These solutions represent the coupled modes of EMPs [25, 26]; the charge $\rho_1$ propagating in the positive direction in the upper channel (Fig. 2) drags small amounts of negative charge $-r\rho_2$, and vice versa. Note that the inter-edge interaction makes the velocity of the coupled modes $\upsilon_{ct} \equiv |\upsilon_j| = \sqrt{U_0^2 - U_{12}^2}$ [$j > 0$ ($j < 0$) for the modes moving to the right (left)] slower than that of the uncoupled modes in the non-interacting regions ($x < 0$ or $x > l$). As an example, let us suppose that a pulse of excess charge is injected from the non-interacting region into the coupled region ($I_{1,\text{in}}$: upper-left in Fig. 2). This excites the right-moving mode ($\rho_1$, $-r\rho_2$) in the coupled region. The remainder of the charge $+r\rho_1$ reflected back into the lower left channel generates the output current $I_{2,\text{out}}$. This process is called "charge fractionalization." These EMP transport characteristics are consistent with the calculations based on the Tomonaga-Luttinger liquid theory [25].

It is convenient to define the EMP scattering matrix $\mathbf{S}_{ct}$ of the coupled transmission lines, which relates the outgoing waves ($I_{1,\text{out}}$, $I_{2,\text{out}}$) to the incoming waves ($I_{1,\text{in}}$, $I_{2,\text{in}}$); see Fig. 2. In terms of the scattering amplitude $r$ at the left ($x = 0$) and right ($x = l$) junctions and the phase evolution $\theta = \omega l/\upsilon_{ct}$ in the coupled region, $\mathbf{S}_{ct}$ is expressed as

$$\mathbf{S}_{ct} = \frac{1}{r^2 - e^{2i\theta}} \begin{pmatrix} (-1+r^2)e^{i\theta} & r(e^{2i\theta}-1) \\ r(e^{2i\theta}-1) & (-1+r^2)e^{i\theta} \end{pmatrix}. \quad (5)$$

The matrix elements are a periodic function of $\theta$ and hence $\omega$ [16]. When $l$ is sufficiently short, such that $\theta \ll 1$, inter-channel coupling can be well approximated by a lumped capacitance $C_X = c_X l$. In this case, the expression for $\mathbf{S}_{ct}$ reduces to

$$\mathbf{S}_{ct} = \frac{1}{\sigma_{xy} + i\omega C_X} \begin{pmatrix} \sigma_{xy} & i\omega C_X \\ i\omega C_X & \sigma_{xy} \end{pmatrix}. \quad (6)$$

When $\omega C_X \ll \sigma_{xy}$, we find $S_{21} \cong i\omega C_X/\sigma_{xy}$. This reproduces the well-known expression for the two-probe admittance of a mesoscopic capacitor, $Y(\omega) = i\omega C_X$ [13-16,20,21].

**C. Quantum point contact**

The EMP transport characteristics through a QPC involve Coulomb coupling between the input and output channels. In this subsection, we derive the scattering amplitudes of a QPC. Figure 3 depicts our model, in which two edge channels are running anti-parallel along each of the split gate metals defining the QPC. We assume that the inter-edge Coulomb interaction exists over a length $l$ along the split gates. As we have shown in Sec. IIB, the effect of the inter-edge Coulomb interaction can be evaluated as the capacitance $c_X$.

The EMP scattering amplitude of the system is calculated by combining the contributions of the electron transport and capacitive transport. When $l$ is much smaller than the wavelength $\lambda_{\text{EMP}}$ of the EMPs, the distributed capacitance can be replaced with the lumped capacitance $C_X = c_X l$. In this case, the output currents $I_{1,\text{out}}$ and $I_{2,\text{out}}$ are given by

$$I_{1,\text{out}} = \sigma_{xy} V_{1,\text{out}} = i\omega C_X (V_{2,\text{in}} - V_{1,\text{out}}) + T_{dc} \sigma_{xy} V_{2,\text{in}} + R_{dc} \sigma_{xy} V_{1,\text{in}},$$
$$I_{2,\text{out}} = \sigma_{xy} V_{2,\text{out}} = i\omega C_X (V_{1,\text{in}} - V_{2,\text{out}}) + T_{dc} \sigma_{xy} V_{1,\text{in}} + R_{dc} \sigma_{xy} V_{2,\text{in}}.$$

Here, $T_{dc}$ and $R_{dc}$ are the dc transmission and reflection probabilities for electron transport, respectively. From these equations, the scattering matrix $\mathbf{S}_{\text{QPC-X}}$ of the QPC is represented as

$$\mathbf{S}_{\text{QPC-X}} = \frac{1}{\sigma_{xy} + i\omega C_X} \begin{pmatrix} R_{dc}\sigma_{xy} & T_{dc}\sigma_{xy} + i\omega C_X \\ T_{dc}\sigma_{xy} + i\omega C_X & R_{dc}\sigma_{xy} \end{pmatrix}. \quad (7)$$

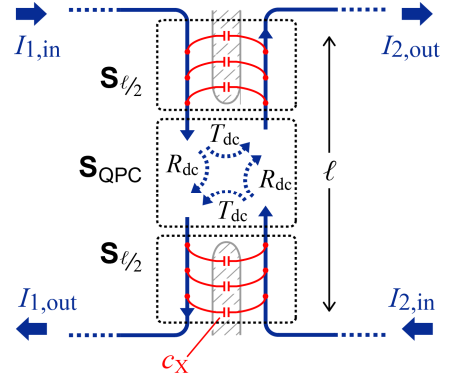

Fig.3 Schematic of edge channels near a QPC. Left and right channels are connected via the QPC. Inter-edge capacitance $c_X$ is distributed over a length $l$ around the QPC.

These scattering amplitudes are controlled by tuning $T_{dc}$. When the QPC is fully open (i.e., $T_{dc} = 1$), the off-diagonal element of $\mathbf{S}_{\text{QPC-X}}$, which we denote by $t_{\text{QPC}}$, is unity. In contrast, when the QPC is pinched off (i.e., $T_{dc} = 0$), EMPs are transferred only through $C_X$, yielding the transmission amplitude $t_{\text{QPC}} = i\omega C_X/(\sigma_{xy} + i\omega C_X)$. Note that, if we set the contribution of Coulomb coupling to zero (i.e., $C_X = 0$) in eq. (7), the diagonal ($r_{\text{QPC}}$) and off-diagonal ($t_{\text{QPC}}$) elements of $\mathbf{S}_{\text{QPC-X}}$ reduce to $R_{dc}$ and $T_{dc}$, respectively.

When $l$ is not sufficiently small to ignore the phase evolution along the interacting region, the lumped-capacitance approximation breaks down, and the distribution of $c_X$ must be taken into account. To calculate $\mathbf{S}_{\text{QPC-X}}$ for the entire system, we define the $\mathbf{S}$ matrices for each section in the system; $\mathbf{S}_{\text{QPC}}$ is the $\mathbf{S}$ matrix for electron scattering at the QPC, and $\mathbf{S}_{l/2}$ is that for Coulomb interaction in coupled channels with length $l/2$ (see Fig. 3). To combine these $\mathbf{S}$ matrices, we convert them to transmission matrices ($\mathbf{T}$ matrices: $\mathbf{T}_{\text{QPC}}$ and $\mathbf{T}_{l/2}$) and calculate the $\mathbf{T}$ matrix of the entire system ($\mathbf{T}_{\text{QPC-X}}$) as [27]

$$\mathbf{T}_{\text{QPC-X}} = \mathbf{T}_{l/2} \mathbf{T}_{\text{QPC}} \mathbf{T}_{l/2}. \quad (8)$$

$\mathbf{S}_{\text{QPC-X}}$ is obtained by reconverting $\mathbf{T}_{\text{QPC-X}}$ to $\mathbf{S}_{\text{QPC-X}}$.

To evaluate $t_{\text{QPC}}$ and $r_{\text{QPC}}$ in a device having a more complicated edge channel geometry in which these channels interact everywhere, the CDE model can be improved as much as necessary by taking the inter-edge capacitances elsewhere into account.

**D. Plasmonic cavity resonators**

A plasmonic cavity is a typical application of EMPs [3-6]. In this subsection, we derive the transmission spectra of single and double plasmonic cavities using the CDE model. We consider closed loops of edge channels having a pair of QPCs at the entrance and exit of the cavity structures. The **S** matrices for individual sections (QPCs and edge channels composing the cavities) are combined to calculate the **S** matrix of the entire system. Here, we consider the lumped capacitance across QPCs ($C_X$) rather than the distributed capacitance ($c_x$) in order to simplify the calculation. For the moment, we ignore EMP dissipation during propagation, assuming $g_{ch} = 0$.

**a) Single plasmonic cavity.** Figure 4(a) shows the schematic of a single-cavity structure. The cavity consists of two QPCs ($QPC_p$: $p$ = L or R) at the entrance and exit. The scattering matrix $\mathbf{S}_p$ of $QPC_p$ is given by eq. (7). On the other hand, the scattering matrix of the looped channel in the cavity, $\mathbf{S}_{Ch}$, can be described by

$$\mathbf{S}_{Ch} = \begin{pmatrix} 0 & \exp(-ik_L L_L) \\ \exp(-ik_U L_U) & 0 \end{pmatrix}.$$

$\mathbf{S}_{Ch}$ represents the phase evolution of the EMPs during propagation. Here, $k_U$ and $k_L$ represent the wave numbers of the EMPs, and $L_U$ and $L_L$ are the lengths of the upper and lower channels in the cavity, respectively. The **S** matrix of the entire cavity ($\mathbf{S}_{Cavity}$) is obtained from the **T** matrix ($\mathbf{T}_{Cavity}$) of the entire system, which is given by

$$\mathbf{T}_{Cavity} = \mathbf{T}_L \mathbf{T}_{Ch} \mathbf{T}_R, \quad (9)$$

where $\mathbf{T}_p$ is the transmission matrix of $QPC_p$ and $\mathbf{T}_{Ch}$ is that of the channel in the cavity; these variables are converted from $\mathbf{S}_p$ and $\mathbf{S}_{Ch}$, respectively. The reflection ($r_{Cavity}$) and transmission ($t_{Cavity}$) amplitudes are the diagonal and off-diagonal terms of $\mathbf{S}_{Cavity}$, respectively.

When the QPCs are pinched off (electron transmission probabilities $T_{dc,L} = T_{dc,R} = 0$), the cavity couples to the input and output channels only through $C_X$. In this case, the transmission spectrum is calculated as

$$|t_{Cavity}| = \left| \frac{-\alpha^2 e^{-if_{ch}\omega/2}}{(1+i\alpha)^2 - e^{-if_{ch}\omega}} \right|, \quad (10)$$

where we assume $L_U = L_L$. In eq. (10), $f_{ch}$ is the normalized frequency $f_{ch} = L/v_{ch}$, where $L = 2L_U = 2L_L$ is the perimeter of the cavity, and $v_{ch} = \sigma_{xy}/c_{ch}$ is the characteristic velocity of the EMPs. Further, $\alpha = 2\pi f_{ch} C_X/\sigma_{xy}$ is the inter-edge coupling between the cavity and the leads at the QPCs.

Figures 4(b) and 4(c) show the calculated transmission spectra as a function of frequency $f$. As shown in Fig. 4(b), $|t_{Cavity}|$ oscillates as a function of the frequency because of the resonance of the EMPs. The fundamental resonance shows a Lorentzian shape in the transmission spectrum, as displayed in Fig. 4(c). With increasing $\alpha$, namely, increasing $C_X$, the resonant peak broadens because the leakage of EMPs increases as $C_X$ increases. In addition, the increase in $\alpha$ reduces the resonant frequency owing to the reduced $v_{EMP}$ in the coupled region near the QPCs.

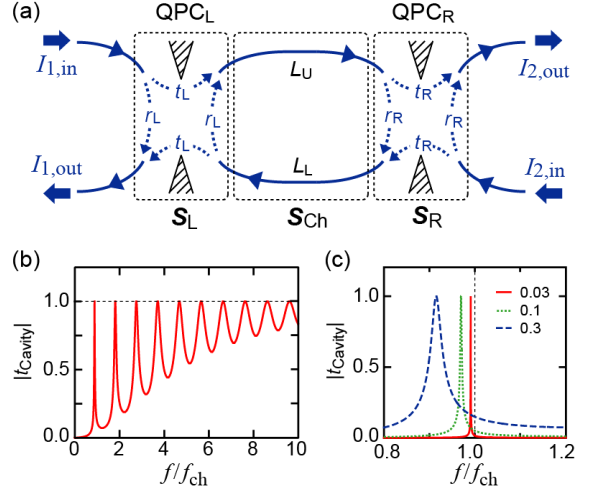

Fig. 4 (a) Schematic of plasmonic cavity confined with $QPC_L$ and $QPC_R$. (b)(c) Calculated $|t_{Cavity}|$ from eq. (10) as a function of normalized frequency $f/f_{ch}$. (b) $|t_{Cavity}|$ in wide frequency band up to $f/f_{ch} = 10$ for $\alpha = 0.03$ and (c) those around the fundamental resonant frequency for $\alpha = 0.03, 0.1,$ and 0.3.

**b) Double plasmonic cavity.** The transmission amplitude $t_{Double}$ of a double-cavity device is derived from the combination of the **S** matrices for different sections. Figure 5(a) shows a schematic of the successive sections of a double-cavity device. The transmission matrix $\mathbf{T}_{Double}$ is obtained from those of the QPCs ($\mathbf{T}_p$: $p$ = L, M, or R) and the channels inside the left and right cavities ($\mathbf{T}_{Ch1}$ and $\mathbf{T}_{Ch2}$) as

$$\mathbf{T}_{Double} = \mathbf{T}_L \mathbf{T}_{Ch1} \mathbf{T}_M \mathbf{T}_{Ch2} \mathbf{T}_R. \quad (11)$$

The scattering matrix $\mathbf{S}_{Double}$ of the double cavity is converted from $\mathbf{T}_{Double}$. The reflection ($r_{Double}$) and transmission ($t_{Double}$) amplitudes are the diagonal and off-diagonal terms of $\mathbf{S}_{Double}$, respectively. Here, we assume the perimeters of the cavities to be $L_{1U} = L_{1L} = L_1/2$ and $L_{2U} = L_{2L} = L_2/2$, where $L_1$ and $L_2$ are the perimeters of the two cavities. As for the single plasmonic cavity, we define the characteristic frequencies $f_{ch1} = v_{ch1}/L_1$ and $f_{ch2} = v_{ch2}/L_2$, where $v_{ch1}$ and $v_{ch2}$ denote the characteristic velocities of EMPs in these two cavities.

As we discussed in Sec. IIC, the transmission amplitude of a QPC involves real (conductive) and imaginary (capacitive) parts. These two types of transmission amplitude produce very different resonance characteristics in a double-cavity device. To elucidate this difference, we consider two cavities coupled via $QPC_M$ without input and output ports, as shown in Fig. 5(b). We describe the currents incoming to and outgoing from $QPC_M$ as $I_{1M,in}$, $I_{2M,in}$ and $I_{1M,out}$, $I_{2M,out}$, respectively. These currents meet the following conditions, where $t_M$ and $r_M$ are the transmission and reflection amplitudes of $QPC_M$.

$$I_{1M,out} = r_M I_{1M,in} + t_M I_{2M,in} = (1-t_M)I_{1M,in} + t_M I_{2M,in} \quad (12)$$
$$I_{2M,out} = t_M I_{1M,in} + r_M I_{2M,in} = t_M I_{1M,in} + (1-t_M)I_{2M,in}.$$

When the resonant frequency of the coupled system is $f_C$, the conditions

$$I_{1M,in} = e^{-2\pi i f_C/f_{ch1}} I_{1M,out},$$
$$I_{2M,in} = e^{-2\pi i f_C/f_{ch2}} I_{2M,out}, \quad (13)$$

are also satisfied. Let us consider the case of $f_{ch1} = f_{ch2}$. One solution of eqs. (12) and (13) is always obtained at $f_C = f_{ch1} = f_{ch2}$ with $I_{1M,out} = I_{2M,out}$. This mode is the in-phase resonance of EMPs in the coupled cavities. In addition, when the coupling $t_M$ is purely capacitive ($t_M$ is an imaginary number), another solution is obtained at $f_C \cong (1 - t_M/i\pi)f_{ch1} = (1 - t_M/i\pi)f_{ch2}$ with $I_{1M,out} = -I_{2M,out}$. This mode is the anti-phase resonance. This anti-phase mode does not appear in conductive coupling, because EMPs with opposite signs cancel each other out at $QPC_M$.

As we discussed in Sec. IIB, the inter-edge Coulomb interaction decreases the velocity of EMPs. For the in-phase resonance, the voltage drop across $QPC_M$ is zero [Fig. 5(c)], so the effect of inter-edge coupling on $v_{EMP}$ disappears. On the other hand, for the anti-phase mode [Fig. 5(d)], the voltage drop across the QPC is doubled, emphasizing the effect of the inter-edge interaction. Consequently, the resonant frequency of the anti-phase mode is always smaller than that of the in-phase mode. In this way, in capacitive coupling, the resonant frequency splits into in-phase and anti-phase modes, whereas only the in-phase resonance is obtained in the case of conductive coupling.

These coupling modes can be seen in simulations using eq. (11). Figure 5(e) displays several simulated transmission spectra of a coupled cavity with $f_{ch1} = f_{ch2}$. Here, we define the coupling parameter $\alpha = 2\pi f_{ch1} C_X / \sigma_{xy}$. When $QPC_M$ is conductive ($T_{dc,M} = 1$, uppermost panel), $|t_{Double}|$ exhibits a single resonant peak near $f/f_{ch1} = 1$ (this corresponds to the second harmonic resonance of the combined cavity). On the other hand, when these cavities are coupled only through $C_X$ ($T_{dc,M} = 0$), the resonant peak splits into two (lower three panels). The splitting $\Delta f$ develops with increasing $\alpha$, namely increasing coupling capacitance $C_X$. This demonstrates that the coupling strength is parameterized with $C_X$ in this double-cavity device. Figure 5(f) shows an image plot of the calculated $|t_{Double}|$ at $T_{dc,M} = 0$ for $\alpha = 0.1$ as a function of the normalized measurement frequency $f/f_{ch1}$ ($x$ axis) and detuning parameter $f_{ch2}/f_{ch1}$ ($y$ axis). A clear anti-crossing of the two resonant modes appears at $f_{ch2}/f_{ch1} = 1$. Thus, the resonance of the EMPs exhibits dramatically different characteristics depending on whether the coupling is conductive or capacitive. Note that the mode splitting in capacitively coupled systems is a key factor in designing plasmonic band structures [11,12].

### III. EXPERIMENTS

In this section, we present the results of EMP transport measurements. First, we compare the measured transmission amplitude $t_{QPC}$ of a QPC with the calculation using

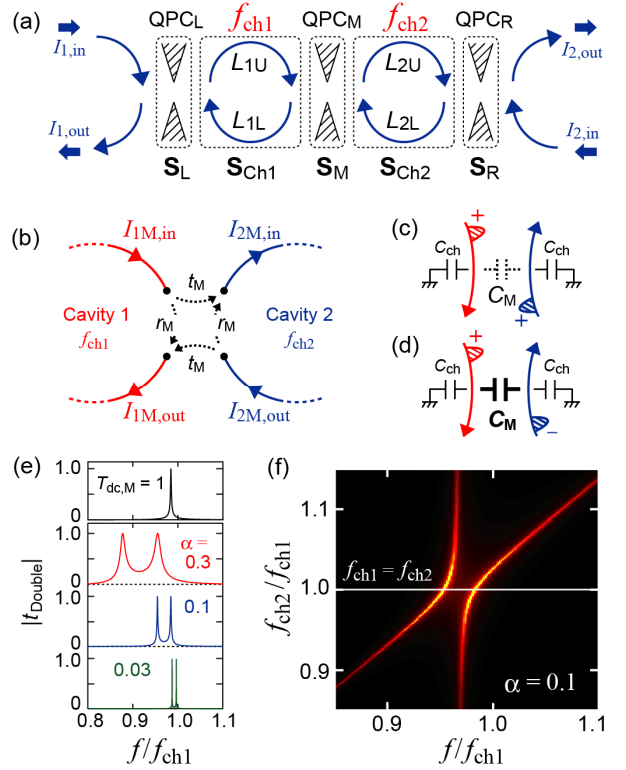

Fig. 5 (a) Schematic of successive sections of a double-cavity structure. (b) Schematic of edge channels near $QPC_M$. (c)(d) Inter-edge coupling with closing $QPC_M$ for dc current. (c) In in-phase coupling, the voltage drop across $QPC_M$ is zero, so $C_M$ can be neglected. (d) In anti-phase coupling, $C_M$ is emphasized. (e) Calculated $|t_{Double}|$ of the coupled cavities at $f_{ch1} = f_{ch2}$ at $T_{dc,M} = 1$ and $\alpha = 0.1$ (upper panel) and at $T_{dc,M} = 0$ and $\alpha = 0.3$, 0.1, and 0.03 (lower three panels). (f) Calculated plot of $|t_{Double}|$ as a function of normalized input frequency $f/f_{ch1}$ ($x$-axis) and $f_{ch2}/f_{ch1}$ ($y$-axis) at $T_{dc,M} = 0$ and $\alpha = 0.1$.

eqs. (5) and (6), in which inter-edge Coulomb interaction across the QPC is evaluated as the inter-edge capacitance. Second, we study the EMP resonance in a gate-defined single-cavity structure. We extract the velocity $v_{EMP}$ and dissipation $g_{ch}$ of EMPs [Fig. 1(b)] in the cavity from the center frequency and $Q$ factor of the resonance. We also show that the resonant frequency, which is a function of $v_{EMP}$, can be tuned with gate voltages applied to a delay gate in the cavity. Third, coupled resonance of EMPs in a double-cavity structure is reported. When the inter-cavity coupling is conductive, we observe a single resonant peak in the transmission spectra, while a double-peak structure appears for purely capacitive coupling, indicative of mode splitting into in-phase and anti-phase resonance.

### A. Device and measurement setup

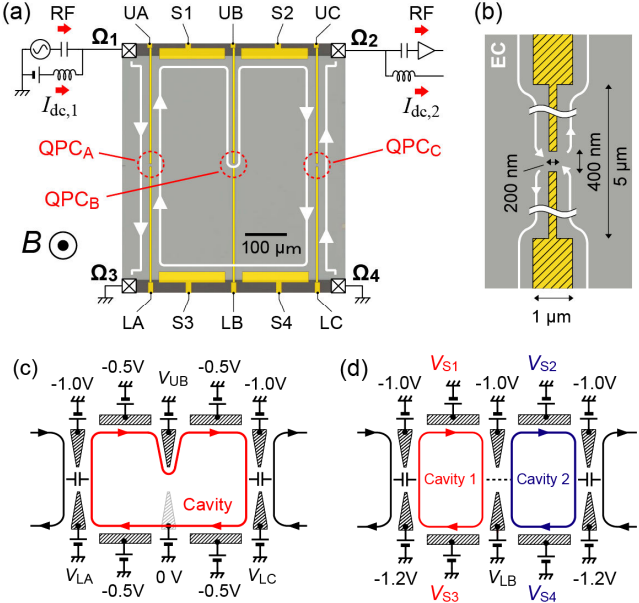
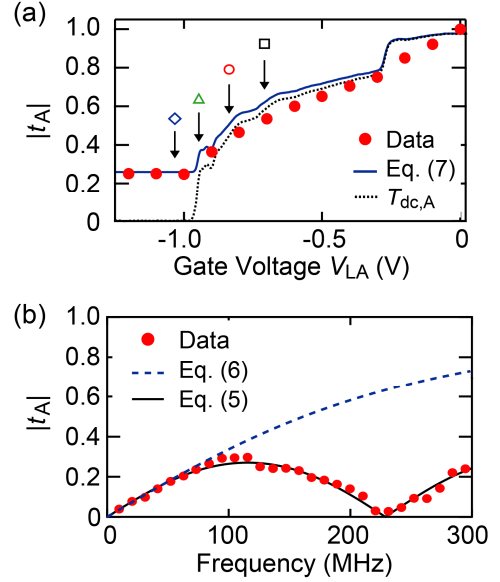

Fig. 6 (a) Colored optical micrograph of device and measurement setup. White lines with arrows show the geometry of the edge channels forming a plasmonic cavity. (b) Schematic of gate metals of QPC$_p$ and edge channels (ECs). (c) Schematic of large single cavity formed with QPC$_A$ and QPC$_C$. (d) Schematic of double cavities formed with all QPCs.

Fig. 7 (a) Transmission amplitude $|t_A|$ of QPC$_A$ as a function of $V_{LA}$ measured at 80 MHz (red circles). Black dotted line is measured $T_{dc,A}$ and blue solid line is simulated $|t_A|$ from eq. (6) assuming $C_X = 22$ fF. Open symbols show the gate voltages, where $|t_{Cavity}|$ was measured as shown in Fig. 10(a). (b) Frequency dependence of the measured $|t_A|$ with QPC$_A$ closed ($T_{dc,A} = 0$ at $V_{LA} = -1.2$ V). Blue dashed line indicates simulation using eq. (6) with $C_X = 22$ fF. Black solid line indicates simulation using eq. (5) with $c_{ch} = 130$ pF/m, $c_X = 47$ pF/m, and $l = 420$ μm.

Figure 6(a) shows a colored optical micrograph of the device and the measurement setup. The EMP devices were prepared in a 2DEG in a GaAs/Al$_{0.3}$Ga$_{0.7}$As heterostructure with electron density $n_e = 1.3 \times 10^{11}$ cm$^{-2}$ and mobility $\mu = 2.1 \times 10^6$ cm$^2$V$^{-1}$s$^{-1}$. They have three split gates [(U$p$, L$p$); $p$ = A, B, or C] to form QPCs and four screening gates (S$q$: $q$ = 1, 2, 3, and 4) to tune the degree of screening for each section of the upper and lower edge channels. The detailed geometry of the gate metals near the QPCs is illustrated schematically in Fig. 6(b). Plasmonic cavities are formed by activating more than one QPC. For example, a large cavity is formed with QPC$_A$ and QPC$_C$ [Fig. 6(c)], while a double cavity is formed with all the QPCs [Fig. 6(d)].

The measurements were performed at 280 mK in a $^3$He cryostat with a high magnetic field (5.3 T) applied perpendicular to the 2DEG (bulk filling factor $\nu = 1$). The chirality of the edge channels was set to clockwise, as shown by the white arrows in Fig. 6(a). In this setup, we applied both dc and rf excitation to the Ohmic contact $\Omega_1$ and simultaneously measured the dc and rf output signals from $\Omega_2$ with the other Ohmic contacts ($\Omega_3$ and $\Omega_4$) grounded. The dc measurements were performed using a standard lock-in technique at 31 Hz.

The electron transmission probability $T_{dc}$ of a gate-defined structure was measured as the collection efficiency $T_{dc} = I_{dc,2}/I_{dc,1}$ of the dc current at $\Omega_2$ for a dc current $I_{dc,1}$ injected from $\Omega_1$. In the $\nu = 1$ QH state, $T_{dc} = 1$ corresponds to $G_{21} = e^2/h$, where $G_{21}$ is the conductance from $\Omega_1$ to $\Omega_2$. We confirmed $T_{dc} > 0.97$ when all the gate voltages were set to zero. This ensures that edge transport is dominant in the present device, and transport through the bulk QH state can be neglected (i.e., $\sigma_{xx} \cong 0$).

The EMP transport properties were investigated by measuring the scattering matrix element $S_{21,meas}$ for rf signals between $\Omega_1$ and $\Omega_2$ using a vector network analyzer. The excitation voltage applied to $\Omega_1$ was about 1.4 mV rms. We estimated the transmission amplitude $t$ of the device as $t = S_{21,meas}/S_{21,meas}^{(0)}$, where $S_{21,meas}^{(0)}$ is the scattering matrix element for full transmission at $V_{LA} = V_{LB} = V_{LC} = 0$ V. With this calibration, the frequency dependence of the measurement system, e.g. the frequency dependence of the insertion loss of the coaxial cables and/or gain of the amplifiers, can be removed.

**B. Quantum point contact**

First, we investigated the electron transmission probability $T_{dc,p}$ through QPC$_p$ ($p$ = A, B or C). We measured

$T_{dc,p}$ as a function of the lower gate voltage ($V_{Lp}$) while applying $V_{Up} = -1.0$ V to the upper gate. The measured $T_{dc,A}$ is shown in Fig. 7(a) by a black dotted line. QPC$_A$ is formed at $V_{LA} \cong -0.25$ V, where $T_{dc,A}$ steeply decreases: $T_{dc,A} < 1$ at this QPC$_A$ definition voltage is a result of the narrow gap of our split gates [400 nm; see Fig. 6(b)], which causes backscattering between the counter-propagating edge channels immediately after the QPC formation. Applying a more negative $V_{LA}$ decreases $T_{dc,A}$ until it reaches $T_{dc,A} = 0$ at around $V_{LA} = -0.95$ V. The small features observed in the range $-0.6$ V $< V_{LA} < -0.95$ V originate from scattering at impurities around the QPC. We observed a similar $V_{Lp}$ dependence of $T_{dc,p}$ for the other two QPCs (not shown).

The absolute transmission amplitude $|t_A|$ of QPC$_A$ measured at 80 MHz is also plotted in Fig. 7(a). Although $T_{dc,A} = 0$ below the pinch-off voltage $V_{LA} \cong -0.95$ V, $|t_A|$ remains finite even after the QPC is fully pinched off because of the Coulomb coupling across the QPC. We modeled these interactions as the inter-edge capacitance across QPC$_A$, as shown in Fig. 3. When QPC$_A$ is closed for dc current, the system can be regarded as coupled 1D transmission lines (Fig. 2). In this case, $|t_A|$ is expected to oscillate as a function of frequency, as discussed in Sec. IIB. Figure 7(b) shows the frequency dependence of the measured $|t_A|$ at $V_{LA} = -1.2$ V (where $T_{dc,A} = 0$). The experimental data are fitted well by eq. (5). From this fitting, the coupling parameters of the present system were found to be $c_{ch} = 130$ pF/m, $c_X = 47$ pF/m, and $l = 420$ μm. These values of $c_X$ and $l$ are comparable to those obtained in our previous study [16].

At low frequencies, the inter-edge Coulomb interaction can be simply described by the lumped capacitance $C_X$ across the QPC. As seen in Fig. 7(b), below 100 MHz, $|t_A|$ monotonously increases with frequency, reflecting the relation $t_A \cong i\omega C_X/\sigma_{xy}$ ($\omega C_X \ll \sigma_{xy}$). Indeed, the data below 100 MHz can be well fitted with eq. (6) with $C_X = 22$ fF. With this parameter, the measured $V_{LA}$ dependence of $|t_A|$ shown in Fig. 7(a) is well reproduced using eq. (7) over the entire range of $V_{LA}$.

**C. Single plasmonic cavity**

In this subsection, we present the results of transmission measurements for a single plasmonic cavity. As sketched in Fig. 6(c), our cavity is defined by using nine gate electrodes. QPC$_A$ and QPC$_C$ are formed by applying a constant gate voltages $V_{UA} = V_{UC} = -1.0$ V to the upper gates UA and UC and varying voltages $V_{LA}$ and $V_{LC}$ to the lower gates LA and LC so that the QPC openings can be tuned. The upper and lower channels in the cavity are defined by the four screening gates, which are biased at $V_{S\beta} = -0.5$ V ($\beta = 1, 2, 3,$ and 4). Additionally, gate UB was exploited to tune the perimeter $L$ of the cavity. EMPs injected from $\Omega_1$ enter the cavity via QPC$_A$ and exit from QPC$_C$ to be collected at $\Omega_2$.

A typical frequency dependence of the transmission amplitude $|t_{Cavity}|$ of the cavity is shown in Fig. 8. The spectrum was obtained with both QPCs pinched off ($T_{dc,A} = T_{dc,C} = 0$ at $V_{LA} = V_{LC} = -1.2$ V). A sharp peak appears near 90 MHz, corresponding to the fundamental resonance. We simulated $|t_{Cavity}|$ using eq. (9), with the inter-edge coupling across the QPCs modeled as a distributed capacitance with $c_{ch} = 130$ pF/m,

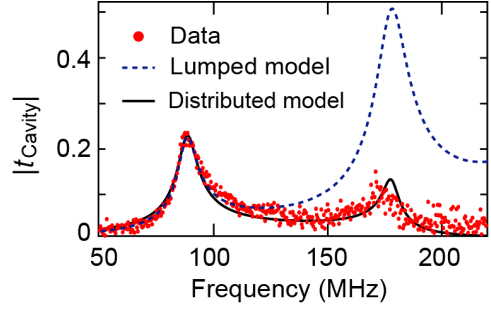

Fig. 8 Typical $|t_{Cavity}|$ of a single plasmonic cavity measured at $V_{LA} = V_{LC} = -1.2$ V and $V_{UB} = -0.36$ V. Black solid line is simulation assuming a distributed capacitance across the QPCs. Blue dashed line shows the fitting assuming lumped capacitance in the frequency band from 50 to 120 MHz.

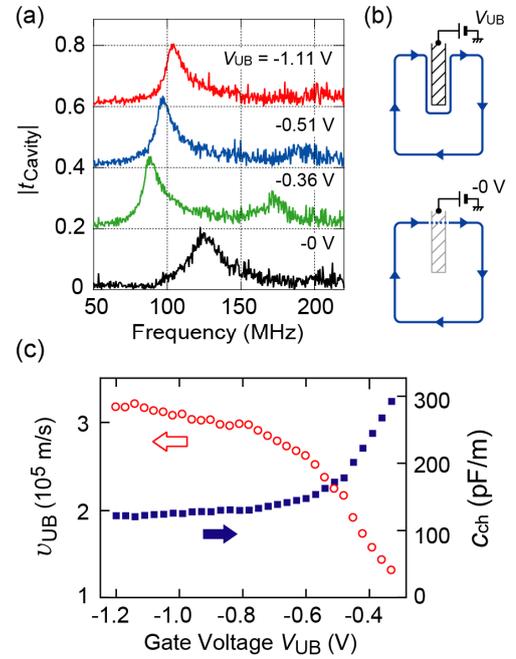

Fig. 9 (a) Transmission spectra $|t_{Cavity}|$ of the single cavity shown in Fig. 6(c) measured at several $V_{UB}$. (b) Geometry of edge channel in the cavity at $V_{UB} = 0$ V (upper) and at $V_{UB} < -0.3$ V (lower). (c) Group velocity $v_{UB}$ (open circles) and channel capacitance $c_{ch}$ (filled squares) along the gate UB derived from the observed resonant frequencies.

$c_X = 47$ pF/m, and $l = 420$ μm. We assumed $g_{ch} = 1/260$ $\Omega^{-1}$m$^{-1}$ to include the dissipation of EMPs; the mean free path of EMPs estimated from this $g_{ch}$ is about 10 mm. The simulation (black solid line in Fig. 8) reproduces the experimental data exceedingly well, including the second harmonic resonance around 180 MHz, indicating the validity of our CDE model.

At low frequencies, the calculation can be further simplified by introducing a lumped capacitance $C_X$ across the QPCs. As shown by the dashed line in Fig. 8, the data below 120 MHz can be well fitted by the lumped capacitance model using the parameters $C_X = 18$ fF, $v_{ch} = 2.17 \times 10^5$ m/s, and $g_{ch} = 1/260$ $\Omega^{-1}$m$^{-1}$. The $C_X$ value is close to that obtained for a single QPC in Sec. IIIB ($C_X = 22$ fF), and the above $v_{ch}$ value is consistent with those previously reported for $\nu = 1$ [5-7,9,10]. In the remainder of this section, we focus on the fundamental resonance and use the lumped capacitance model to simplify the calculation.

Figure 9(a) shows the transmission spectra measured at several values of $V_{UB}$ with both QPCs pinched off ($T_{dc,A} = T_{dc,C} = 0$). When $V_{UB}$ is changed from 0 V to −0.36 V, the resonant frequency $f_{RES}$ shifts from 125 to 85 MHz. At $V_{UB} = 0$ V, EMPs take a shortcut under the gate UB, which makes a short cavity perimeter of $L \cong 1750$ μm, whereas at $V_{UB} < -0.3$ V, EMPs make a detour, and $L$ increases to $L \cong 2250$ μm [Fig. 9(b)]. As $V_{UB}$ decreases further to more negative values, $f_{RES}$ gradually increases and saturates at $f_{RES} \cong 105$ MHz at $V_{UC} < -0.8$ V. This is a consequence of the gate voltage dependence of the velocity $v_{UB}$ along gate UB [9]. We extracted the velocity $v_{UB}$ by comparing the resonant frequencies for the cases with and without a detour around UB. The obtained $v_{UB}$ values are shown in Fig. 9(c) as a function of $V_{UB}$. As $V_{UB}$ decreases from −0.3 to −1.2 V, $v_{UB}$ increases by a factor of three. This suggests that, by appropriately designing the cavity structure, it is possible to vary $f_{RES}$ electrostatically over a wide range by up to a factor of three. In Fig. 9(c), we also plotted the channel capacitance $c_{ch}$ along UB deduced from the $v_{UB}$ values using eq. (2). As $V_{UB}$ decreases, $c_{ch}$ decreases, reflecting the growing distance between the channel and the gate, which decreases $c_m$ between them. The extracted value of $c_{ch}$ (about a few hundred pF/m) is close to the value of $c_m$ estimated from a finite element calculation (a few hundred pF/m, assuming a distance of a few micrometers between the channel and the gate metal and a channel width of a few hundred nanometers [28]).

Figure 10(a) shows the effect of QPC openings on $|t_{Cavity}|$. The four spectra shown were taken at different gate voltages $V_L \equiv V_{LA} = V_{LC}$ ranging from −1.02 to −0.70 V [marked by open symbols in Fig. 7(a)] corresponding to different values of $T_{dc}$ ($\equiv T_{dc,A} = T_{dc,C}$). The data show that the resonant peak becomes sharper with decreasing $V_L$ and hence decreasing $T_{dc}$. At $V_L = -1.02$ V, where both QPCs are pinched off and the coupling is purely capacitive, the peak height becomes $|t_{Cavity}| \cong 0.2$ with $Q = f_{RES}/\Delta f_{FWHM} \cong 7$, where $\Delta f_{FWHM}$ is the full width at half maximum (FWHM) of the resonant peak. As shown in Fig. 10(b), the measured $Q$ factor increases with decreasing $T_{dc}$. This is consistent with the intuitive expectation that the resonance is sharper when the leakage of EMPs through the QPCs is smallest. Calculation of the $Q$ factor using eq. (9) reveals that $g_{ch} = 1/260$ $\Omega^{-1}$m$^{-1}$ provides a good account of the obtained $T_{dc}$ dependence of $Q$ [solid line in Fig. 10(b)]. If we set $g_{ch} = 0$ in eq. (9), the resultant $Q$ factor deviates from the data at low $T_{dc}$, reaching $Q > 20$ at $T_{dc} = 0$ [dashed line in Fig. 10(b)]. This means that when the QPCs are pinched off, $Q$ is limited not by the leakage at the QPCs, but by the dissipation of EMPs during propagation,

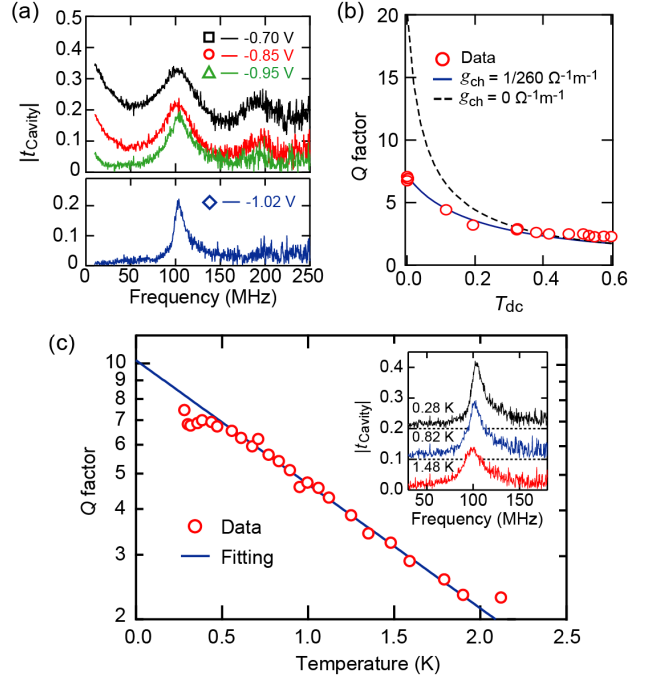

Fig. 10 (a) Measured $|t_{Cavity}|$ at several $T_{dc}$ [$V_L = -0.7, -0.85, -0.95$, and $-1.02$ V; marked by open symbols in Fig. 7(a)] at $V_{UB} = -1.2$ V. (b) Quality factor of the measured resonant peaks (circles) as a function of $T_{dc}$ with those calculated from eq. (9) for $g_{ch} = 1/260$ $\Omega^{-1}$m$^{-1}$ (solid blue line) and for $g_{ch} = 0$ (dashed black line). (c) (Main panel) Temperature dependence of $Q$ factor of the fundamental resonance in the single cavity. (Inset) Transmission spectra of the resonance obtained at various temperatures.

namely, by $g_{ch}$. From these results, we can expect that a higher $Q$ factor is obtained in a smaller cavity.

To explore the origin of EMP dissipation, we measured the temperature dependence of the $Q$ factor. The inset of Fig. 10(c) shows transmission spectra obtained at several temperatures. The resonant peak is most pronounced at the lowest temperature, and the $Q$ factor monotonically decreases with increasing temperature [Fig. 10 (main panel)]. The temperature dependence of $Q$ is fitted well by an exponential function, $Q_0 \exp(-T_s/T_{res})$, where $Q_0$ is the $Q$ factor at a temperature of 0 K, and $T_s$ is the temperature of the system. $T_{res}$ represents the characteristic temperature of the resonance. We found $T_{res} = 1.4$ K from this fitting, which is comparable to the Zeeman gap ($\cong 1.2$ K, assuming the $g$-factor in the GaAs/Al$_{0.3}$Ga$_{0.7}$As heterostructure to be $|g| \cong 0.4$) at $B = 5.3$ T. This result may suggest that excitation of down-spin charge carriers gives rise to the dissipation of EMPs. Further detailed experiments, such as space- and/or spin-resolved EMP measurements, would reveal the dominant mechanism of EMP dissipation.

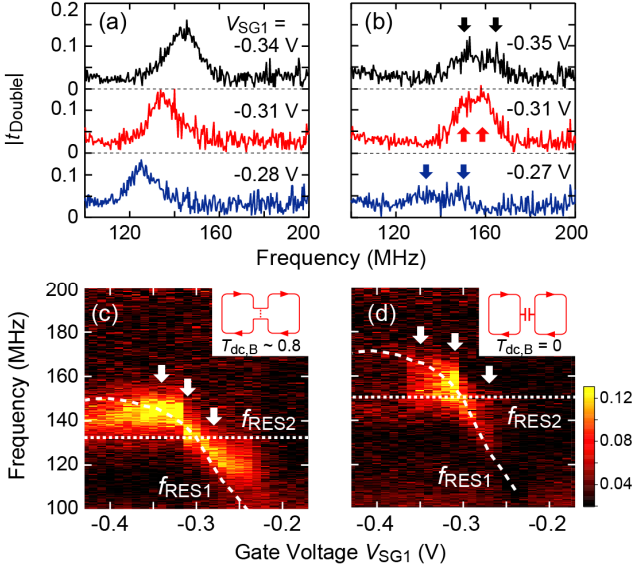
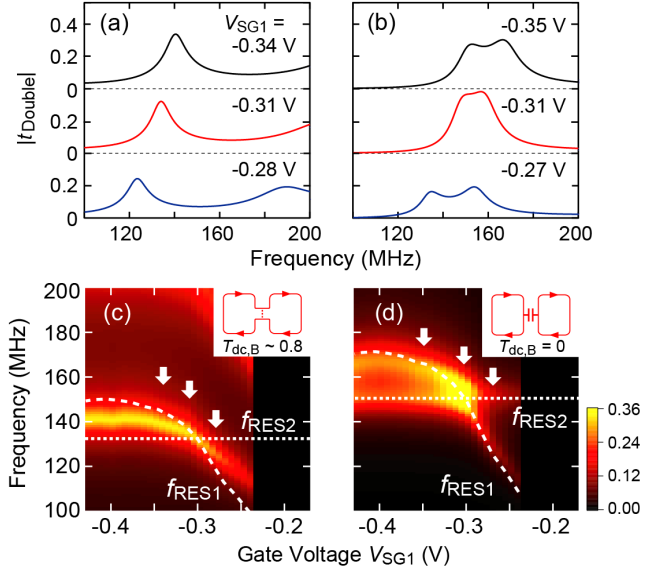

Fig. 11 Measured transmission spectra of the coupled cavities obtained at various $V_{SG1}$ with $V_{SG2}$ fixed at (a) $T_{dc,B} \cong 0.8$ and (b) $T_{dc,B} = 0$. Arrows in (b) indicate the spectral peaks. Image plots of $|t_{Double}|$ measured at (c) $T_{dc,B} \cong 0.8$ and (d) $T_{dc,B} = 0$. Insets in (c) and (d) show the schematics of the edge channels in the double-cavity structure.

Fig. 12 (a)(b) Calculated transmission spectra and (c)(d) image plots of $|t_{Double}|$ of the coupled cavities under the same conditions as in Fig. 11.

### D. Coupled plasmonic cavities

In this subsection, we examine the coupling of the resonance in a double-cavity structure. As shown in Fig. 6(d), the double cavity is defined by activating gate LB as well as UB, and thus by forming $QPC_B$, which separates the entire cavity into left cavity (cavity 1) and right cavity (cavity 2) components. We applied fixed voltages of −1.0 V to the upper gates (UA, UB, and UC) and −1.2 V to gates LA and LC, but varied $V_{LB}$ of gate LB to tune the transmission amplitude $t_B$ of $QPC_B$. The resonant frequencies of cavity 1 ($f_{RES1}$) and 2 ($f_{RES2}$) can be independently controlled by tuning $V_{SG1} \equiv V_{S1} = V_{S3}$ and $V_{SG2} \equiv V_{S2} = V_{S4}$, respectively. These cavities have the same perimeters of $L_1 = L_2 = 1350$ μm. We measured the absolute transmission amplitude $|t_{Double}|$ of the double-cavity device while changing $V_{SG1}$ and fixing $V_{SG2} = -0.3$ V.

When $QPC_B$ is open ($T_{dc,B} \cong 0.8$) at $V_{LB} = -0.3$ V, EMP transport through $QPC_B$ is almost conductive ($|t_B| \cong T_{dc,B}$), as shown in Fig. 7(a). Figure 11(a) shows typical resonant spectra of the double cavity in this conductive-coupling regime, taken at several $V_{SG1}$ values. Figure 11(c) displays the image plot of $|t_{Double}|$ as a function of $V_{SG1}$ ($x$ axis) and the measurement frequency ($y$ axis). To highlight the effect of inter-cavity coupling, the intrinsic resonant frequencies $f_{RES1}$ of cavity 1 and $f_{RES2}$ of cavity 2 are indicated by the dashed and dotted lines, respectively, superposed on the image plot: $f_{RES1}$ is measured in the absence of cavity 2 with $V_{LC} = 0$ V, and vice versa. As expected, $f_{RES1}$ varies with $V_{SG1}$ in a manner similar to that seen in Fig. 9(c), whereas $f_{RES2}$ is independent of $V_{SG1}$. Consequently, $f_{RES1}$ and $f_{RES2}$ cross each other as a function of $V_{SG1}$. The spectra shown in Fig. 11(a) were taken near the crossing point, which is indicated by the arrows in Fig. 11(c). As seen in Figs. 11(a) and (c), in the conductive-coupling regime, only a single resonant peak is observed, whose resonant frequency roughly lies between $f_{RES1}$ and $f_{RES2}$.

On the other hand, when $QPC_B$ is pinched off ($T_{dc,B} = 0$) at $V_{LB} = -1.2$ V, the coupling between the two cavities is purely capacitive and is mediated by inter-edge Coulomb interaction. Figures 11(b) and (d) show similar measurements in this capacitive-coupling regime. The spectra taken near the crossing of $f_{RES1}$ and $f_{RES2}$, shown in Fig. 11(b), reveal a double-peak structure, which is pronounced when $f_{RES1} \neq f_{RES2}$ [highlighted by the black and blue arrows in Fig. 11(b)]. Even at $f_{RES1} \cong f_{RES2}$, an additional shoulder peak structure is visible (shown by the red arrows), suggesting a splitting of the resonant frequency [Figs. 5(e) and (f)].

We simulated the resonance characteristics of the double cavities using eq. (11) and compared these with the experimental results. Figure 12 shows the results of this calculation, where we assumed the lumped inter-edge capacitances $C_p$ across $QPC_p$ to be $C_A = C_C = 18$ fF, $C_B = 10$ fF, and the EMP dissipation to be $g_{ch} = 260$ $\Omega^{-1}$m$^{-1}$. These calculations match the measured results well; at $T_{dc,B} \cong 0.8$, a single resonant peak appears between $f_{RES1}$ and $f_{RES2}$, whereas at $T_{dc,B} = 0$, double-peak structures are seen in the spectra. At $f_{RES1} = f_{RES2}$ ($V_{SG1} = -0.31$ V) and $T_{dc,B} = 0$, a shoulder peak is also reproduced in the simulation.

This excellent agreement between the experiment and the simulation demonstrates the validity of our CDE model. It also provides evidence for the crossover from conductive to capacitive inter-cavity coupling controlled by $V_{LB}$. At $f_{RES1} = f_{RES2}$, the single resonant peak observed for the conductive coupling corresponds to the in-phase mode, while the double peak (or shoulder peak) observed for the capacitive coupling indicates both the in-phase and anti-phase modes. Although the size of the mode splitting in the capacitively coupled cavities cannot be quantified in the present device, clear anti-crossing of two resonant frequencies [Fig. 5(f)] will be observed by optimizing the cavity structures. Furthermore, we can expect to realize plasmonic crystals [11,12] using capacitively coupled multiple cavities. Our CDE model would assist in the design of such an integrated plasmonic device.

## IV. DISCUSSION

In our CDE model, Coulomb interactions are represented as distributed electrochemical capacitances. As discussed in detail in Sec. II, the transport characteristics of plasmonic devices can be calculated by considering capacitive couplings between the edge channel and the other conductors present in the surroundings. This approximation allows us to compute the EMP transport conveniently even in integrated circuits comprising multiple plasmonic devices. However, because the screening effect in low-dimensional electronic systems is not sufficient to fully screen the long-range Coulomb interactions, in real devices, essentially all parts of the QH device can be coupled to each other [16]. In an integrated system, the complicated networks of edge channels may make it unfeasible to evaluate the **S** matrices for individual sections. Partial screening of the edge channels by appropriately designed metal structures would solve this problem. The striking sensitivity of the EMP velocity to the presence of nearby metals [9] indicates that such screening of edge channels is indeed active; this suggests the possibility of limiting the spatial range of Coulomb coupling by using metals.

To design integrated plasmonic circuits, the coherence of EMPs is a key requirement. Although we have assumed that $g_{ch}$ is a frequency-independent constant, it is not clear whether this is the case. The dissipation mechanism has been studied previously both theoretically and experimentally. Volkov and Mikhailov assumed the loss of EMP charge into the bulk region and found that the damping is proportional to the frequency [1]. This corresponds to the frequency-independent $g_{ch}$ in our circuit representation. On the other hand, Talyanskii *et al.* found the damping to be proportional to $f^2$ in their experiment [6]. This frequency dependence implies $g_{ch} \propto f$ in our model. The temperature dependence of the $Q$ factor [Fig. 10(c)] may suggest that the former mechanism is dominant in our device. However, the dissipation mechanism has not yet been completely clarified. We expect that gate-defined cavities like those used in this study would provide a means of solving this problem, allowing us to evaluate the frequency dependence of the $Q$ factor, and hence of the EMP dissipation.

## V. CONCLUSION

A CDE circuit model has been developed to describe the EMP transport properties in a QH device. A characteristic impedance and distributed electrochemical capacitances were introduced as fundamental physical quantities that characterize the edge channels constituting plasmonic circuits. The EMP transport characteristics of various gate-defined structures, such as a QPC and single and coupled cavities, were calculated using this model. Good agreement was found between the calculations and the experimental results.

The advantages of our CDE model can be summarized as follows. First, it allows the effect of Coulomb interactions to be expressed as capacitances, i.e., as circuit parameters. This provides a simple and powerful means of evaluating the impact of Coulomb interactions in electronic systems, which is essential to understanding electron dynamics in QH edge channels. Because the capacitance between two conductors can be measured experimentally or simulated using, e.g., finite element methods, the EMP transport properties can be computed using realistic parameters obtained separately. Second, the circuit representation allows us to analyze EMP transport in multiple plasmonic devices using the scattering matrix formalism. This opens a way to establish future integrated plasmonic circuits.

## ACKNOWLEDGMENTS

We appreciate the experimental support given by M. Ueki. This study was supported by the Grants-in-Aid for Scientific Research (21000004 and 21810006) and the Global Center of Excellence Program through the "Nanoscience and Quantum Physics" project of the Tokyo Institute of Technology from the Ministry of Education, Culture, Sports, Science and Technology of Japan (MEXT).